\documentclass[%
 aip,
 amsmath,amssymb,
reprint, twocolumn 
]{revtex4-1}

\usepackage{graphicx}
\usepackage{dcolumn}
\usepackage{bm}

\usepackage[utf8]{inputenc}
\usepackage[T1]{fontenc}
\usepackage{mathptmx}
\usepackage{etoolbox}
\usepackage[separate-uncertainty=true]{siunitx}

\makeatletter
\def\@email#1#2{%
 \endgroup
 \patchcmd{\titleblock@produce}
  {\frontmatter@RRAPformat}
  {\frontmatter@RRAPformat{\produce@RRAP{*#1\href{mailto:#2}{#2}}}\frontmatter@RRAPformat}
  {}{}
}%
\makeatother
\begin{document}

\preprint{AIP/123-QED}

\title[Optical Bias and Cryogenic Laser Readout of a Multipixel SNSPD]{Optical Bias and Cryogenic Laser Readout of a Multipixel Superconducting Nanowire Single Photon Detector}
\author{Frederik Thiele}
 \affiliation{Institute for Photonic Quantum Systems (PhoQS), Paderborn University, Warburger Str. 100, Paderborn, Germany}
  \affiliation{Department of Physics, Paderborn University, Warburger Str. 100, Paderborn, Germany}
\email{frederik.thiele@uni-paderborn.de}
\author{Niklas Lamberty}%
  \affiliation{Department of Physics, Paderborn University, Warburger Str. 100, Paderborn, Germany}
\author{Thomas Hummel}
\affiliation{Institute for Photonic Quantum Systems (PhoQS), Paderborn University, Warburger Str. 100, Paderborn, Germany}
\author{Tim Bartley}
\affiliation{Institute for Photonic Quantum Systems (PhoQS), Paderborn University, Warburger Str. 100, Paderborn, Germany}
  \affiliation{Department of Physics, Paderborn University, Warburger Str. 100, Paderborn, Germany}

\date{\today}

\begin{abstract}
Cryogenic opto-electronic interconnects are gaining increasing interest as a means to control and read out cryogenic electronic components. The challenge is to achieve sufficient signal integrity with low heat load processing. In this context, we demonstrate the opto-electronic bias and readout of a commercial four-pixel superconducting nanowire single-photon detector array using a cryogenic photodiode and laser. We show that this approach has a similar system detection efficiency to a conventional bias. Furthermore, multi-pixel detection events are faithfully converted between the optical and electrical domain, which allows reliable extraction of amplitude multiplexed photon statistics. Our device has a passive heat dissipation of \SI{2.6}{mW}, maintains the signal rise time of \SI{3}{ns}, and operates in free-running (self-resetting) mode at a repetition rate of \SI{600}{kHz}. This demonstrates the potential of high-bandwidth, low noise, and low heat load opto-electronic interconnects for scalable cryogenic signal processing and transmission. 
\end{abstract}

\maketitle

\section{Introduction} \label{sec:intro}

Many emerging quantum technologies, from superconducting quantum circuits to superconducting single-photon detectors, require cryogenic conditions in which they operate~\cite{Wendin2007,Natarajan2012c,Devoret2013,Krinner2019,Castillo2023}. As the size of these systems increases, a plurality of interconnects are required in order to address each component reliably~\cite{Krinner2019}. Nevertheless, developing scalable interfaces between room-temperature control and cryogenic operating environments remains an on-going challenge. 

Typically, coaxial cables are used to address and read out such cryogenic circuitry. However they may be limited by noise pickup, heat load, and signal processing fidelity, particularly in the RF-domain~\cite{Krinner2019,McCaughan2018}. Recently, cryogenic opto-electronic interconnects are being explored to address these issues~\cite{Youssefi2021,Sahu2022,Thiele2023b}. Working in the optical domain offers the potential of high bandwidth, no electrical interference, and low heat load. Crucially, these interconnects need to faithfully convert between electrical and optical signals under cryogenic conditions. On the optical to electronic input, this requires cryogenic photodiodes, and for the electronic to optical output, cryogenic modulators or lasers. 
 \begin{figure}[h]
    \centering
    \includegraphics[width=0.9\linewidth]{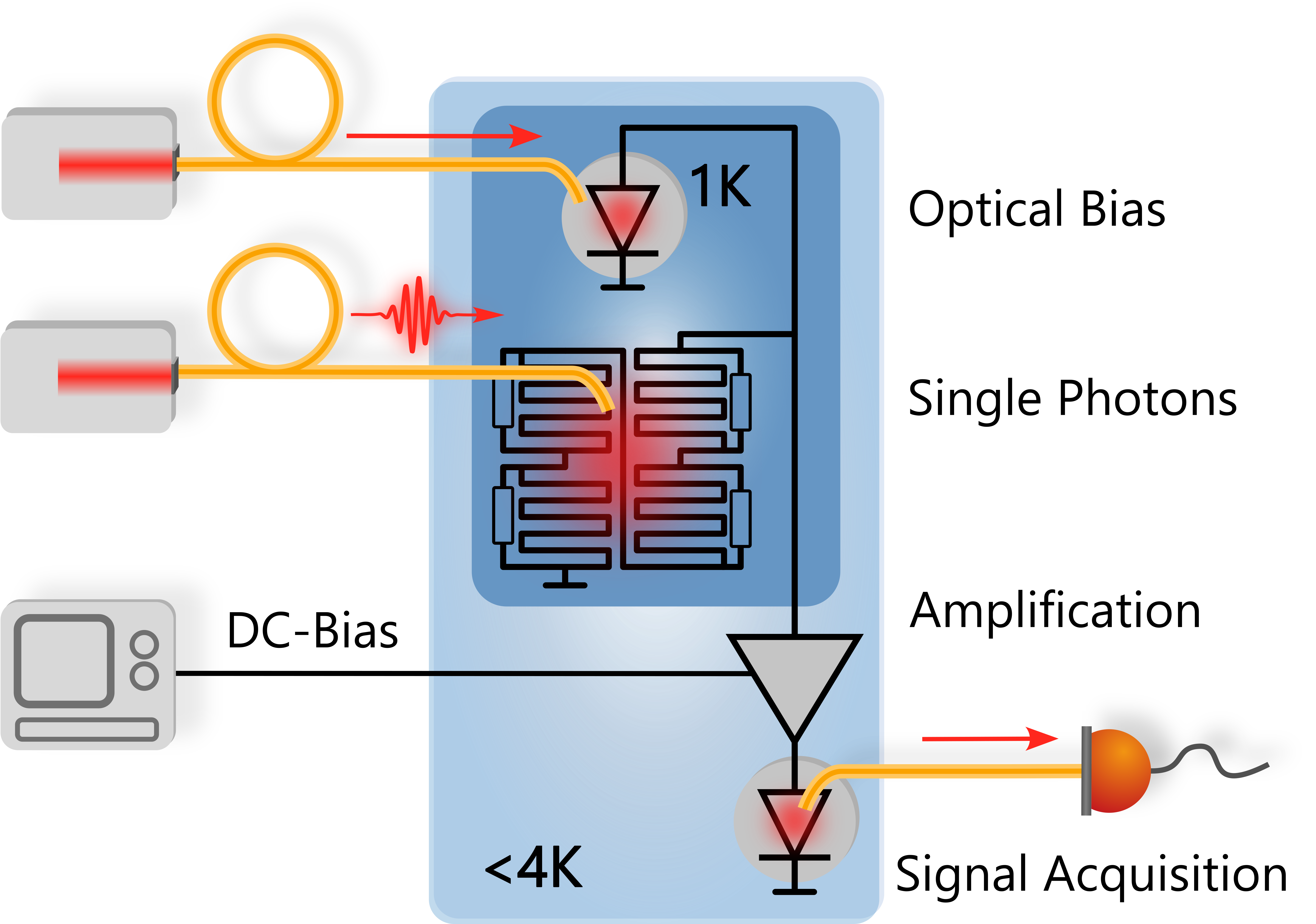}
    \caption{Schematic layout for the opto-electronic operation of the SNSPD. When the cryogenic photodiode is illuminated, a current is generated to bias the SNSPD. Detection signals from the four pixel SNSPD are transmitted through the amplifier to modulate the intensity of a cryogenic laser diode. The resulting optical signal is read out at room temperature.}
    \label{fig:Overview}
\end{figure}
 \begin{figure*}
    \centering
    \includegraphics[width=0.9\textwidth]{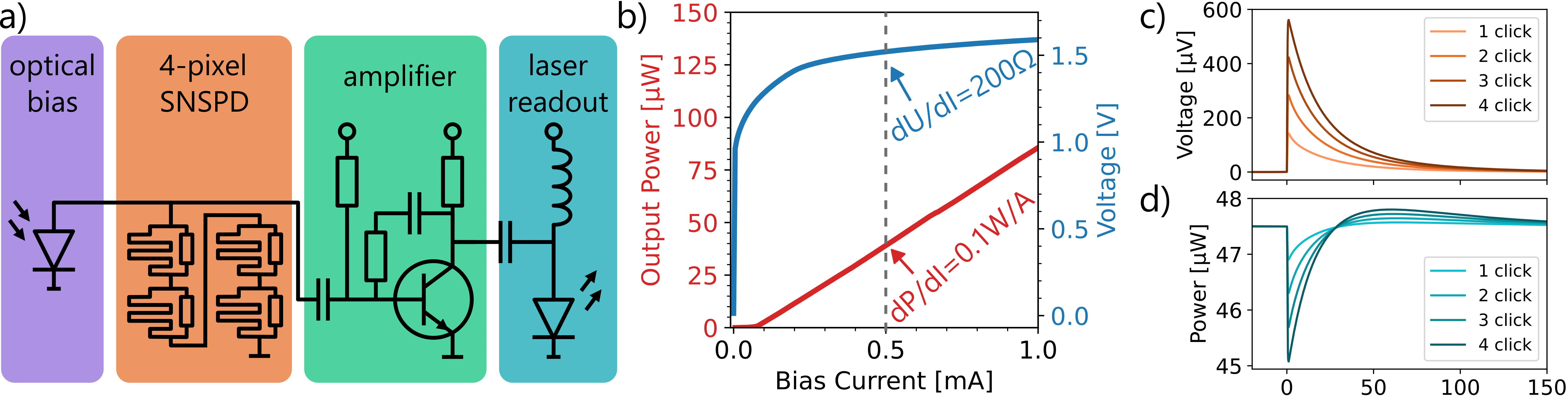}
    \caption{a) Electrical layout for the opto-electronic operation of an SNSPD.  In this layout the four pixel SNSPD is shown and can be exchanged with a single pixel SNSPD. The bias for the transistor and laser diode is provided by DC-current sources at room temperature. b) Acquired IV-curve and IP-curve of the laser diode below \SI{4}{K}. c) Simulated detection pulses of the four-pixel SNSPD. The voltage trace is acquired at the output of the four-pixel SNSPD. d) Converted detection signal (simulated) into an optical modulation at the laser diode .}
    \label{fig:ElecLayout}
\end{figure*}

Such cryogenic opto-electronic interconnects have been used to control or read out superconducting circuits in a variety of configurations. Initially Nakahara et al.~\cite{Nakahara1994} has shown an optical input with a photo detector and readout of a Josephson Junction with a laser diode. The laser readout of superconducting circuits was further investigated with high bandwidth laser diodes by Fu et al.~\cite{Fu2022a}. To provide fast current pulses to Josephson Junctions or other single flux circuits, photodiodes have been used as a optical to electrical converter~\cite{Morse1994pd,Shinada2010,Kieler2019,Lecocq2021}. To read out these cryogenic circuits optically, electro-optic modulators have been used~\cite{Youssefi2021,Sahu2022,Shen2024}.

More recently, this approach has been applied to bias~\cite{Thiele2022c} and read out~\cite{DeCea2020,Thiele2023b} superconducting nanowire single photon detectors (SNSPDs). These detectors have become an indispensable tool for low-light level applications due to their near unity detection efficiencies~\cite{Chang2021,Reddy2020a}, high timing resolution~\cite{Korzh}, low dark counts~\cite{Hochberg2019} and integration potential~\cite{Ferrari2018a}. A major frontier of this technology is increasing scale: i.e. large arrays of many detectors~\cite{Wollman2019,Oripov2023}, multiple waveguide integrated SNSPDs~\cite{Cheng2023}, and photonic quantum computing~\cite{Deng2023}. A major challenge when operating large numbers of SNSPDs is scalability of electrical connections for bias and readout.

In this paper, we demonstrate an opto-electronic operation of an SNSPD by using a photodiode for the biasing and a cryogenic laser for readout. We apply this approach to a multi-pixel SNSPD, whereby the output signal is multiplexed on a single optical readout line. The fidelity of the electronic to optical conversion is sufficient to decode the number of pixels which click from the amplitude of the optical readout signal. In contrast to previous approaches for optical readout~\cite{Thiele2023b,DeCea2020}, we simultaneously operate in the free-running (self-reset) regime and maintain the rise- and fall times, compared to conventional electronic bias and read out. The latent power consumption of this device is in the range of \SI{2.6}{mW}.

\section{Bias and readout scheme} \label{sec:methods}
Operating an SNSPD requires connections between the room-temperature control and acquisition electronics, and the device itself operating in a cryogenic environment. A thin-film superconducting nanowire is current-biased below its critical current to maintain superconductivity. When a single photon is absorbed by the wire, superconductivity breaks down, forming a resistive hotspot and changing the wire's impedance. The resulting impedance change generates the SNSPD's detection signal, observed as a voltage pulse. In order to reset back to the superconductive state, the current can be diverted through a resistor in parallel to the superconducting wire. By connecting multiple SNSPDs ("pixels") in series, we can register multiple detection events when photons impinge on different pixels simultaneously. The height of the voltage pulse therefore becomes the sum of the voltages produced by each pixel individually.

In our opto-electronic approach, the SNSPD's bias current is supplied by a cryogenic photodiode. To read out the detection signal of the SNSPD, we amplify the detection signal with a cryogenic amplifier and use the amplified electronic signal to modulate the power of an output laser, as shown in Fig.~\ref{fig:ElecLayout}~a).  

Previously, we demonstrated the operation of an SNSPD with a cryogenic photodiode without significant performance degradation~\cite{Thiele2022c}, as well as in the context of all-optical operation~\cite{Thiele2023b}. The use of a cryogenic photodiode offers an advantage over conventional current sources, notably in reducing heat load and facilitating the generation of fast current pulses for optical gating of the SNSPD.

An opto-electronic readout of the SNSPD's detection signal requires a conversion of the detector's impedance change into an optical modulation. A laser is a suitable candidate to convert a resulting current change into an optical modulation, since laser diodes offer a high responsivity and modulation bandwidth even at cryogenic temperatures \cite{Nakahara1994,Bardalen2018,Wu2021}. Furthermore, we require low-loss fiber coupling for an efficient signal transmission, in particular at cryogenic temperatures. We characterized the output power and IV-characteristics of an off-the-shelf pigtailed laser diode below \SI{4}{K}, as shown in Fig \ref{fig:ElecLayout}~b). To do so, we sweep the bias current and measure the operation voltage and output power. The threshold current of the laser diode is \SI{77}{\micro A}. The diode is packaged with a single-mode fiber pigtail and emits light around \SI{1550}{nm}. The responsivity of the diode, including the fiber coupling, is \SI{0.095\pm0.01}{W/A}. In principle, this responsivity could enable a direct readout of an SNSPD where the laser acts as the electronic load. In practice, minor bias instabilities can lead to detector latching. We address this challenge by incorporating a low-power cryogenic amplifier to isolate laser biasing noise from the SNSPD and enhance the signal amplitude. 

The low-noise amplifier circuit is based on a silicon germanium heterojunction bipolar transistor (SiGe-HBT). In this configuration, we DC-bias the transistor's base and collector to optimize impedance matching and minimize power consumption. We introduce negative feedback to stabilize the amplifier with additional capacitors and resistors between the base and collector. Furthermore, we chose a bias tee for the laser diode with a RF-transmission above \SI{10}{MHz} for the SNSPD's pulses. An S-parameter analysis of the amplifier circuit is provided in the Supplementary Material. 

\section{Electronic modeling}
The characterization of the cryogenic photodiode, amplifier and laser diode generates a set of parameters which enables the electronic behavior to be modeled. With the addition of an SNSPD simulation model~\cite{Berggren2018a}, we can simulate the complete opto-electronic readout circuit. To do so, we approximated the behavior of the cryogenic photodiode with a photovoltaic diode model~\cite{Kennerud1969,Cubas2014,NDetigma2018} and the data set of Thiele et al.~\cite{Thiele2022c}. Based on the characterization of the cryogenic amplifier, we approximated the SiGe-transistors with a room temperature Gummel-Poon-model~\cite{Gummel1970,infineonBFP640} which is adjusted for our operation temperature. The cryogenic laser diode is approximated using the Shockley diode model~\cite{Shockley1949} with a numerical approximation~\cite{Yapo2024}. Finally, we simulate the multi-pixel SNSPD using four identical single-pixel SNSPDs given by the SPICE model~\cite{Berggren2018a}, and adjust the inductance to approximate the real devices. 

In the simulation, we operate the four-pixel SNSPD at \SI{17}{\micro A} which is provided by the photodiode. To demonstrate the amplitude-multiplexed readout of this SNSPD, we generate four simulations with an increasing number of individual pixels triggered. The generated detection signal at the detector is displayed in Fig.~\ref{fig:ElecLayout}~c). The inverting amplifier and laser diode transform the signal into an optical modulation as shown in Fig.\ref{fig:ElecLayout}~d). The laser's operation point is chosen such that the highest gain of \SI{0.095\pm0.01}{W/A} is reached, compare Fig.~\ref{fig:ElecLayout}~b). The intensity modulation of the photodiode reaches an amplitude of \SI{0.6}{\micro W} which can be detected with conventional photodiodes. 

\section{Setup and results}
\subsection{Experimental setup} \label{sec:ExpSetup}
To operate the SNSPD opto-electronically, we placed the circuits into a cryostat. The SNSPD and photodiode were operated below \SI{1}{K} on a single stage. Due to the cooling power limitations of this stage, the amplifier and laser diode are placed on a cooling stage operated at \SI{3}{K}. The circuits are interconnected by short coaxial cables. The SNSPD was biased by illuminating the photodiode with a laser at \SI{1530}{nm}, of which the power was changed by a variable attenuator. The amplifier and laser diode bias are generated by room temperature electrical power supplies.
\begin{figure}[h]
    \centering
    \includegraphics[width=1\linewidth]{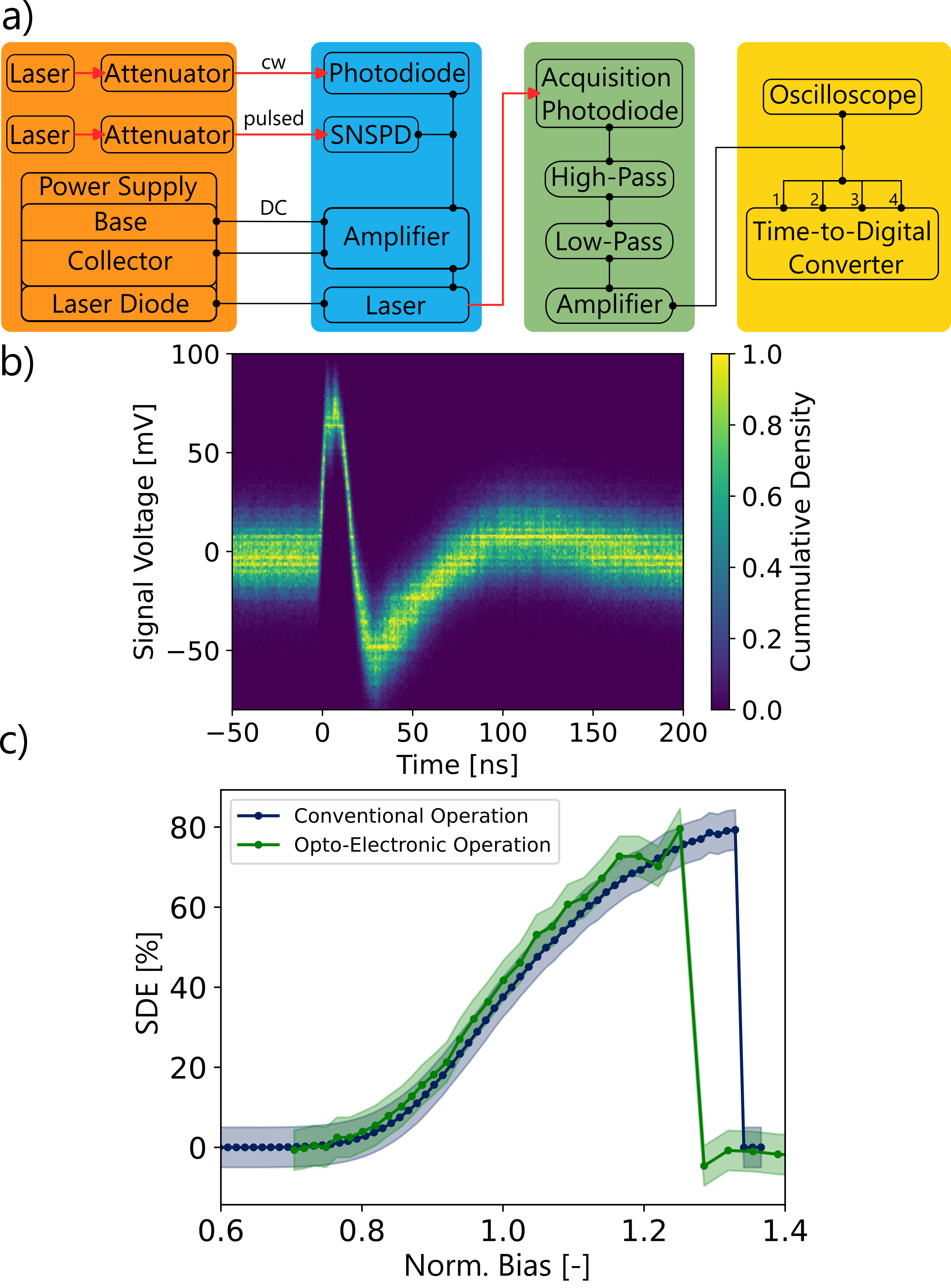}
    \caption{a) Schematic experimental setup of the opto-electronic SNSPD operation. The operation power for the amplifier and laser diode is supplied to the cryogenic circuit by an external power supply. To trigger the SNSPD, laser pulses are generated which are attenuated down to the single photon level. The setup can be operated with the single SNSPD or with the four-pixel SNSPD. The detection signals are read out with an amplified InGaAs-photodiode. The resulting output signal is further filtered with electrical filters until the signal is acquired with an oscilloscope and time-to-digital converter. b) Detection signal of the single-pixel SNSPD converted by the laser diode and acquired with an amplified photodiode. We acquired 3000 traces and generated in every time step a histogram of the voltages. The heat map shows the cumulative density of the single photon traces. c) Comparison of the SNSPD's system detection efficiency (SDE). The bias power is normalized to the point of inflection in the SDE. The same SNSPD is operated in the conventional operation and in the opto-electronic operation.}
    \label{fig:SingleSNSPD}
\end{figure}

For an efficient optical to current conversion, we choose an InGaAs-InP photodiode from Marktech (MTPD1346D-010) which was pigtailed and achieved a responsivity of \SI{0.65}{A/W}~\cite{Thiele2022c}. We use two types of SNSPDs in the opto-electronic operation. At first we tested as a single pixel WSi-SNSPD with a critical current of \SI{5.6}{\micro A} in the conventional operation~\cite{Thiele2022c}. Furthermore, we operated a commercial four-pixel SNSPD (Photonspot), which has a critical current of \SI{17}{\micro A} and was previously characterized by Tiedau et al.~\cite{Tiedau2020}. The intermediate cryogenic amplifier is based on a SiGe-transistor (Infineon BFP640). An off-the-shelf pigtailed laser diode from Thorlabs (LPS-1550-FC) facilitates the optical signal conversion. Details on the amplifier and laser diode characterization are presented in the supplementary material~\ref{sec:SUP}. 

The optical readout signals are the transmitted by single mode optical fiber and acquired by an amplified photodiode at room temperature, as shown in Fig.~\ref{fig:SingleSNSPD}~a). The resulting electronic signals are acquired by an oscilloscope (Teledyne Waverunner 9254) and shown in the supplementary materials \ref{sec:SUP}. An optimal readout performance of the single pixel was reached with an SFP-module (Finisar FWLF-1519-7D-59) as a photodiode and bandwidth filtering with an additional low-pass filter (Mini circuits VLF-45+ (\SI{45}{MHz})). To maintain a good signal separation with the four pixel-SNSPD, we used an amplified photodiode (Thorlabs), a low pass filter (Mini circuits BLP-25+ (25MHz)) as well as a second amplifier (Minicircuits ZKL-2R5+ (10MHz-2500MHz)). To analyze the click statistics we include an additional time-to-digital converter (TDC) (Swabian Instruments Time Tagger Ultra). In the four pixel readout, we split the detection signal into four channels to acquire four measurement outcomes simultaneously, see Fig.\ref{fig:SingleSNSPD}~a) and \ref{fig:4Pix}~a). 

In the single- and four-pixel SNSPD characterizations we extract a photon statistic from a coherent input state. To do so, we generate \SI{9}{ps}-laser pulses using a diode laser at \SI{1556}{nm} (Alphalas PICOPOWER-LD-1550-50-FC) and attenuate the mean photon number down to the single photon level. As a reference, we calibrate the input state with an SNSPD with a known detection efficiency of $83\pm5\%$. To generate a constant time reference with a \SI{600}{kHz} repetition rate, we use synchronization pulses from the pulsed laser as a trigger for the oscilloscope and TDC. 

\subsection{Single SNSPD} \label{sec:SingleSNSPD}
To test our approach, we first compare the performance of a single-pixel SNSPD using conventional and opto-electronic operation. The opto-electronic readout signal is acquired by an SFP module and recorded on an oscilloscope; a collection of 3000 signal traces of the detection events are shown in figure~\ref{fig:SingleSNSPD}~a). The detection signal has a rise time of \SI{3}{ns} and fall time of about \SI{100}{ns}. A clear over-swing of the trace is visible, generated by the bandwidth filtering of the bias tees in the amplifier circuit. To reduce the electronic noise, we added low pass filters in the room temperature readout. Furthermore, the SNSPD trace has a signal-to-noise ratio of 5.3 (compared by the noise-RMS), which is limited by high frequency noise. In summary, the SNSPD showed a clear single photon response, self-resets after the detection and maintained a fast rise time.

The system detection efficiency (SDE) can be determined by illuminating the SNSPD with a calibrated input described in section~\ref{sec:ExpSetup}. The mean photon number $\bar{n}$ can then be extracted from the photon count rate ($PCR$), dark count rate ($DCR$), the repetition rate ($RR$) of the laser and the detection efficiency of the detector $\eta$; $\bar{n}=log(1-(PCR-DCR)/RR)\eta^{-1}$ \cite{Schapeler2020}. The SDE of the SNSPD is then extracted by illumination at the same input power level and than illuminating the detector under test. To compare the operation methods, we sweep the bias of the SNSPD and extract the SDE. Under conventional electrical bias, the bias current is varied by an external power supply. For the opto-electronic bias, we vary the input optical power to the cryogenic photodiode.
To compare the bias levels of the two different approaches fairly, the bias is normalized to the point of inflection in the SDE, which typically follows a sigmoidal curve. This analysis is required since we cannot measure the current \emph{in situ}, and is described in detail in Ref.~\cite{Thiele2022c}.

The conventional and opto-electronic method reach a comparable SDE, see Fig.~\ref{fig:SingleSNSPD}~c). The opto-electronic operation does not reach the full bias level due to premature latching which is likely induced by electrical noise leaking into the SNSPD bias from the laser diode. In summary, the SDE reaches in the conventional method $79.3\pm5\%$ and the opto-electronic operation $79.5\pm5\%$. The dark count rate of the opto-electronic bias was elevated from \SI{1}{kcps}s to \SI{30}{kcps} due to the light scattering of the photodiode and laser diode. This scattering can be reduced by improved shielding of the components in the cryostat.

\subsection{Photon Number Resolution with a four-pixel SNSPD} \label{sec:4Pix}
\begin{figure}
    \centering
    \includegraphics[width=0.95\linewidth]{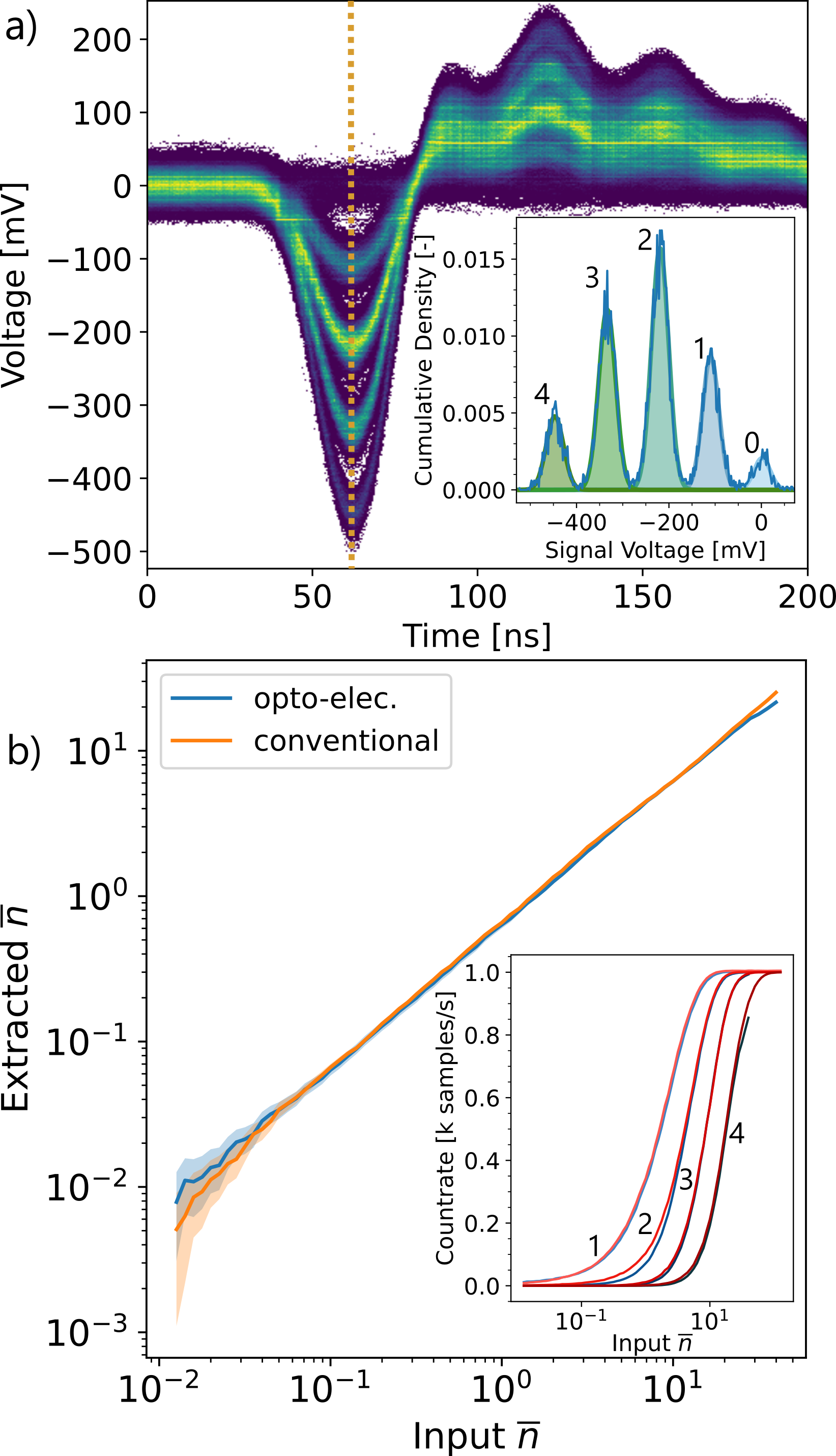}
    \caption{a) Readout signal acquired from the four-pixel SNSPD. The laser's output signal is acquired with an amplified photodiode and oscilloscope. The traces are acquired with an effective \SI{10}{MHz} to \SI{25}{MHz} bandpass filter. The inset shows the voltage distribution of the traces at a \SI{62}{ns} delay. The histogram is normalized to the total number of compared traces (9900). b) The extracted mean photon number $\bar{n}$ of the four pixel SNSPD. The input photon number is varied with a variable attenuator. The reference photon number is extracted with a reference SNSPD with a detection efficiency of $83\pm5\%$. The inset shows the count rate acquired by the TDC. The thresholds of the TDC are set between Gaussian curves as seen in the inset of Fig.~a)}
    \label{fig:4Pix}
\end{figure}

Building on the opto-electronic readout of a single SNSPD, we want to exploit the linear gain of the opto-electronic laser with a multi pixel SNSPD. This array consists of four SNSPDs biased in series with a \SI{50}{\ohm} resistor parallel to each detector~\cite{Tiedau2020}. The voltage pulses generated by each pixel add up on the readout line and thus encode the measured photon-number in the total pulse height which we simulated in section \ref{sec:methods} and Fig.~\ref{fig:ElecLayout}~d) and d). The detector array is coupled to a single optical fiber and thus acts as a spatially multiplexed detector with a quasi photon number resolution. The opto-electronic readout needs to maintain a clear amplitude separation by a proportional gain in the cryogenic amplifier and laser diode. As a result, variation in the amplitude of the detectors is converted into an output power change in the laser diode. 

The resulting detection traces with distinct amplitude levels are acquired with an oscilloscope, as shown in Fig.~\ref{fig:4Pix}~a). A clear the detection levels can be extracted in the cumulative overlay of the traces as shown in the inset in Fig.~\ref{fig:4Pix}~a). 

In order to verify the operation of the array as a quasi photon number resolving detector, we expose the detector array to pulsed coherent states with differing mean photon numbers. The real time readout of the photon number is achieved by splitting the output of the photodiode on four channels of a TDC. The mean photon number is scanned using a variable optical attenuator. The measured photon number distribution has the dark count rate of each photon number subtracted and is fitted with a Poissonian distribution and the previously known positive operator-valued measure (POVM) matrix of the detector~\cite{Schapeler2020}. As shown in Fig.~\ref{fig:4Pix}~b). The POVM matrix attributes probabilities for measuring different outcomes based on the photon number in the measured state. The fitted mean photon numbers are highly linear with attenuation and in good agreement with the fitted results from the conventional operation. The array's primary feature of photon number resolution is thus fully maintained with optical readout.

\section{Conclusion} \label{sec:conclusion}
In summary, we have successfully demonstrated that cryogenic opto-electronic interfaces can faithfully reproduce signals from a four-pixel SNSPD. By modulating the output intensity of a fiber-coupled cryogenic laser, we convert the electronic detector signal to the optical domain. Furthermore, the linear gain of the amplifier allows us to faithfully recover the number of pixels which simultaneously click, encoded in the amplitude of the laser light. This is achieved with a latent power dissipation of \SI{2.6}{mW}. Along with the high bandwidth capacity of optical signals, this promises high potential for scalability.

Nevertheless, improvements can be made to the laser, in particular optimizing devices with a low threshold current and improving high frequency power stability at \SI{4}{K}. Furthermore, (hybrid) integration of all components would allow for more compact devices with increased electronic performance. This approach also complements other readout schemes such as row-column~\cite{Wollman2019} and frequency modulation~\cite{Sypkens2024}. More generally, the amplification stage acts as a tunable impedance matching buffer, which can be adapted to drive other components directly from the output of an SNSPD.

\section{Supplementary Material} \label{sec:SUP}
Additional information on the electronic circuit and characterization is given in the supplementary material. We provide for the information about the IV-curve of the photodiode and laser diode, as well as a characterization of the amplifier. The timing jitter and the raw traces of the used SNSPDs given as well.

\section{Data Availability}
We provide openly the datasets of the IVP laser diode characterization, the acquired SNSPD traces, the acquired countrates and the SPICE models used in this publication (https://doi.org/10.5281/zenodo.10818631). 

\section{Acknowledgements} \label{sec:Ack}
This work was supported by the Bundesministerium für Bildung und Forschung (Grant No. 13N14911) and (Grant No. 13N15856). We thank Varun Verma (NIST) for providing the superconducting films for the Superconducting Nanowire Single Photon Detectors, Hermann Kahle for the discussion on cryogenic laser diodes and Timon Schapeler for the POVM-matrix of the four-pixel SNSPD.

\nocite{*}
\bibliography{Bibo,Bibo2}

\section{Supplementary Material}

\section{Electronic characterization}
The electronic modeling of the cryogenic circuit requires an initial dataset for the comparison. To do so, we characterized the photodiode, amplifier and laser diode individually at cryogenic temperatures. In the following, we show the characterization and adapted models electronic components. The complete SPICE model including the SNSPD model can be used as a LTSPICE model. (See for further information URL).

\subsection{Amplifier}
\begin{figure}[h]
    \centering
    \includegraphics[width=0.9\linewidth]{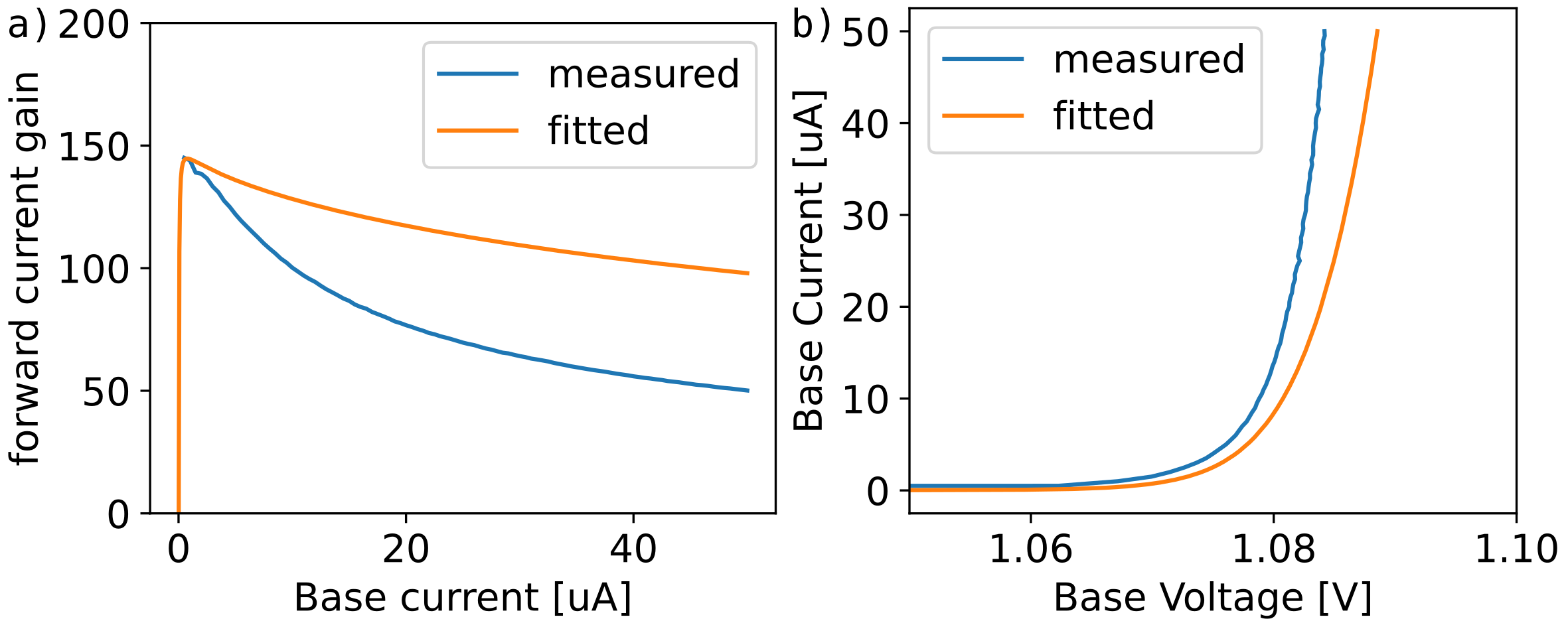}
    \caption{a) The forward current gain $h_{fe}$ of the BFP640 at \SI{4}{K}. b) The diode curve of the base emitter junction at \SI{4}{K}. The fitting was done by manually adjusting the values of the room temperature Gummel-Poon model, provided by Infineon.}
    \label{fig:transistor_spice}
\end{figure}

\begin{figure}[h]
    \centering
    \includegraphics[width=0.9\linewidth]{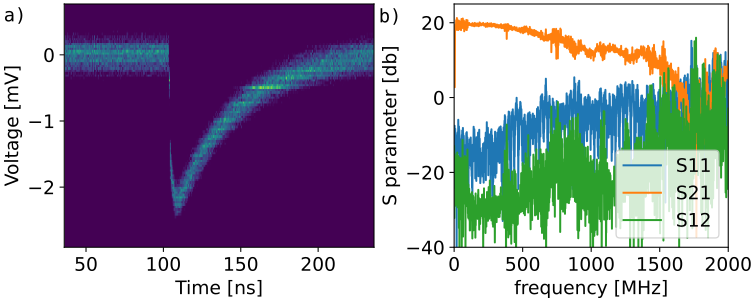}
    \caption{a) The detection signal of the single pixel SNSPD after the amplifier with conventional readout. b) the S-parameters of the amplifier. S11 is the reflection coefficient, S21 is the amplification and S12 is the reverse Isolation of the device. The rise of the S11 and S12 parameters towards higher frequencies is due to low transmission through the cryogenic RF-lines and the noise floor of the VNA used here.}
    \label{fig:amplifier_characterisation}
\end{figure}

For initial design of the amplifier, the transistor has been characterized separately and a Gummel-Poon model has been fitted to the measured values as can be seen in Fig \ref{fig:transistor_spice}. The Gummel-Poon model does not account for all relevant transport mechanisms at \SI{4}{K}, which requires unrealistic values for some parameters to achieve a decent fit.
In order to ensure proper operation, the amplifier circuit has been characterized separatly from the optical readout. For this we use a Vector Network analyzer (Picotech PicoVNA-108) with a \SI{20}{dB}  attenuator on the amplifier input to achieve an input power level of \SI{-40}{dBm} which is similar to that expected from an SNSPD. The measured S-parameters can be seen in Fig.~\ref{fig:amplifier_characterisation} (b) The amplifier is also tested with an SNSPD, where it maintains free running operation and a low distortion output signal. The output trace without a laserdiode as readout can be seen in Fig. \ref{fig:amplifier_characterisation}~b).

\subsection{Cryogenic photodiode}
\begin{figure}[h]
    \centering
    \includegraphics[width=0.9\linewidth]{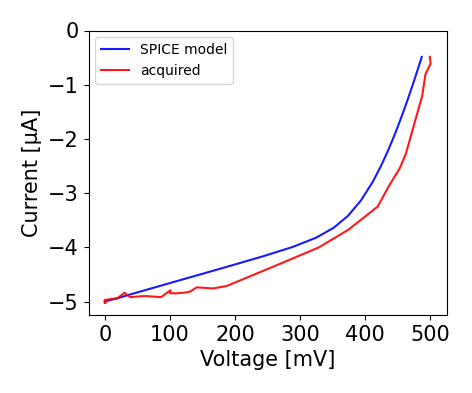}
    \caption{Characterization of the cryogenic photodiode. The photodiode is illuminated under a constant input power of \SI{7.8}{\micro A}. An electronic load to the photodiode is varied while the voltage and current is acquired. In addition, a SPICE model is approximated for the photodiode. }
    \label{fig:IVModelPD}
\end{figure}

The photodiode was characterized below \SI{4}{K}. To do so, we illuminated the diode and vaied the serial load at room temperature. Simultaneously, we measured the output voltage in parallel and measured the generated current in series, see Fig.~\ref{fig:IVModelPD}. Using this dataset, we approximate the photodiode with SPICE model \cite{Kennerud1969,Cubas2014,NDetigma2018} as shown in Fig.~\ref{fig:IVModelPD}. The model SPICE model is described by a set of subcircuit model (CITE URL).    

\subsection{Cryogenic laser diode}
\begin{figure}[h]
    \centering
    \includegraphics[width=0.9\linewidth]{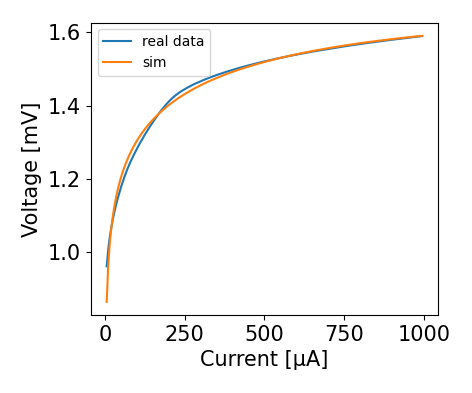}
    \caption{Electronic characterization of the cryogenic laser diode. The diode is operated below \SI{4}{K}. The input current is varied while the voltage of the diode is measured. An additional SPICE model for the diode is approximated based on the Shockley model.}
    \label{fig:IVLaserdiode}
\end{figure}

The laser diode is characterized by applying a bias current and acquiring the output power and voltage across the diode. We approximate the diode by a Shockley model~\cite{Shockley1949} which we can convert to a SPICE model~\cite{Yapo2024}. Figure~\ref{fig:IVLaserdiode} displays the IV-characterization of the diode and the approximated SPICE model.

\section{List of Materials}
\begin{figure}[h]
    \centering
    \includegraphics[width=\linewidth]{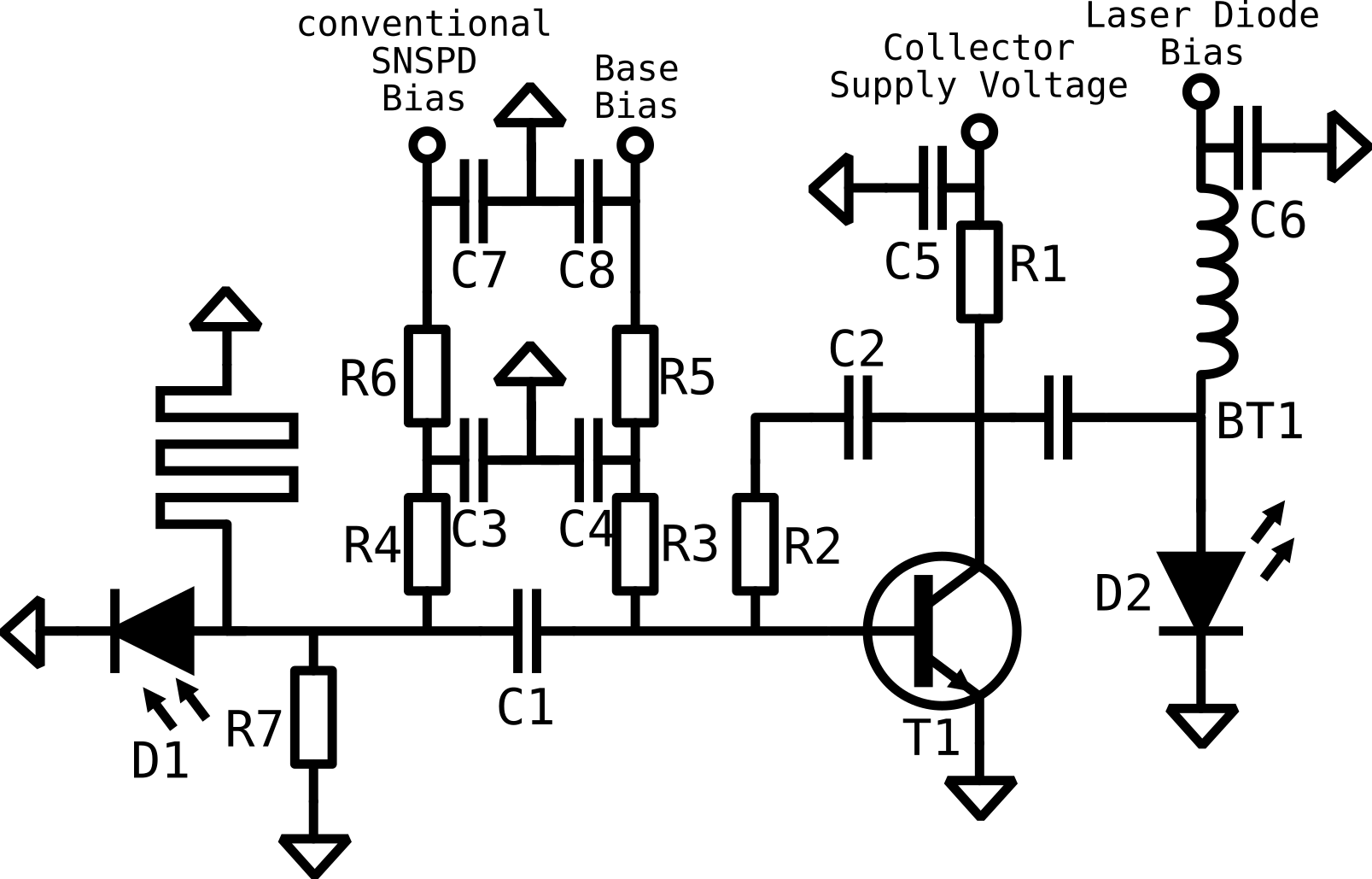}
    \caption{The full electrical circuit diagram of the electronics used for the opto-electronic readout. The circuit is the same for the 4 pixel device except for resistor R7 as the load resistors are integrated with the 4 pixel device.}
    \label{fig:full_laser_readout}
\end{figure}

\begin{figure}
    \centering
    \includegraphics[width=1\linewidth]{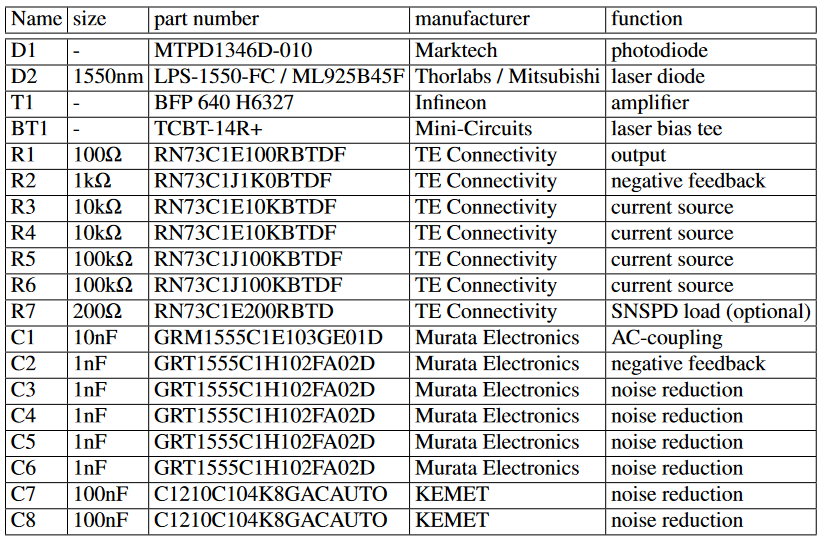}
    \caption{The components used in the electrical circuit as shown Fig. \ref{fig:full_laser_readout}.}
    \label{tab:billofmaterial}
\end{figure}

The opto-electronic readout of the SNSPD is schematically presented in the main manuscript. We display in Figure~\ref{fig:full_laser_readout} the complete circuit diagram including the biasing with ground-to-DC-Voltage capacitors for noise isolation. Furthermore, we added a DC-biasing channel to operate the SNSPD independent to the photodiode bias of the SNSPD for testing purposes. All component used in the circuits are off-the-shelf components, as displayed in the table~\ref{tab:billofmaterial}. 

\section{Heat Load Estimation}
\begin{figure}
    \centering
    \includegraphics[width=1\linewidth]{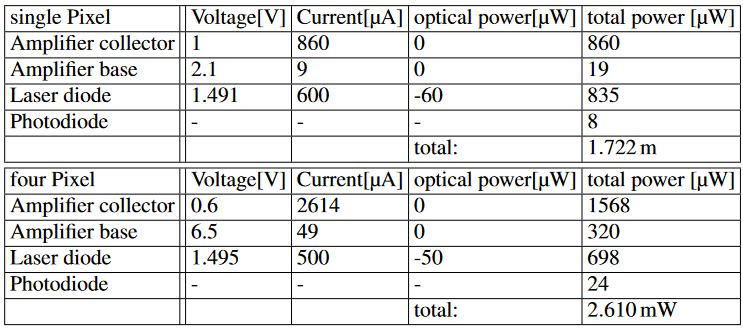}
    \caption{The approximate power dissipation of all components used in the readout scheme not including heatload through coaxial cables. All components except the photodiode are located in the \SI{4}{K} Stage of the cryostat.}
    \label{tab:heatload}
\end{figure}

The electronic components dissipate heat in the cryostat due to the need to bias them. We can estimate the heat load of the various components by their operation voltage and current. We used slightly different operating points for the single-pixel and four-pixel SNSPD, which is accounted for in table~\ref{tab:heatload}. This analysis does not take the heatload from the cabling into account. 

\section{Jitter}

The jitter of the single pixel SNSPD in the convetnional and opto-electronic operation as a comparison. To do so, the SNSPD is operated at the same bias point as in the countrate comparison. As an input trigger, laser pulses with a width of \SI{9}{ps} are transmitted onto the SNSPD at \SI{1550}{nm}. The detection signals are then detected with a time-to-digital converter. In the opto-electronic operation the electronic filters are added in the jitter measurement as described in the section IV of the main manuscript. The jitter of the SNSPD has a full width half maximum (FWHM) in conventional operation of \SI{489}{ps} and a FWHM of \SI{1380}{ps} in the opto-electronic operation. We relate the increase in the jitter in the opto-electronic operation to the higher electronic noise level. The noise in the laser readout could be reduced by a different laser diode or by a stronger electronic filtering.

\begin{figure}[h]
    \centering
    \includegraphics[width=0.9\linewidth]{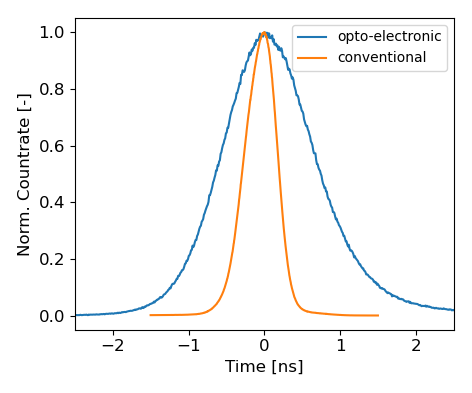}
    \caption{Jitter measurement of the single Pixel SNSPD. The jitter is acquired for the same SNSPD has a FWHM in conventional operation of \SI{489}{ps} and a FWHM of \SI{1380}{ps} in the opto-electronic operation.}
    \label{fig:enter-label}
\end{figure}

\section{Raw Signal Traces}
\begin{figure}[h]
    \centering
    \includegraphics[width=0.9\linewidth]{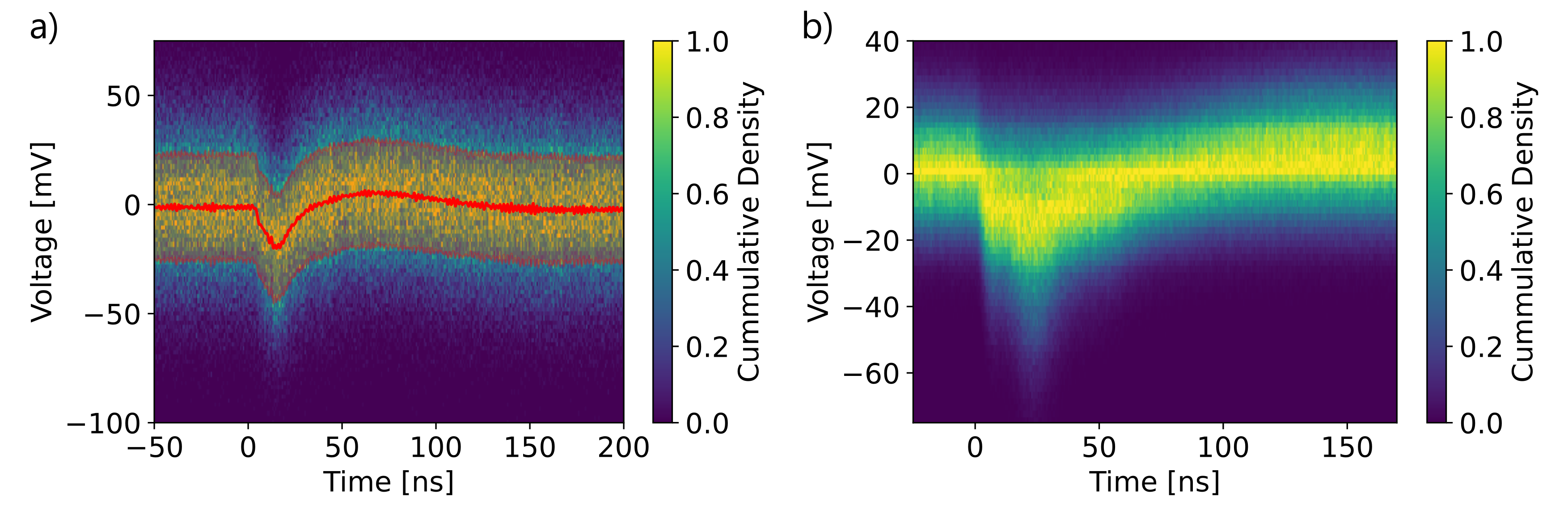}
    \caption{a) Cumulative signal trace of the single-pixel SNSPD. The traces are acquired with a photodiode and oscilloscope and without additional filters. In addition, the average signal (red) is presented with a band (orange) of the width of the standard deviation. b) Cumulative traces acquired with the four-pixel SNSPD.}
    \label{fig:TracesSUP}
\end{figure}

The signal traces of the opto-electronic operation of the SNSPD are acquired as described in the main manuscript in the section experimental setup. Figure~\ref{fig:TracesSUP} displays the signal traces directly from the acquisition photodiode and without the additional electronic filters. In the acquisition of the single pixel SNSPD the photodiode of an SFP-module is used, as shown in Fig.~\ref{fig:TracesSUP}~a). The four-pixel SNSPD is read out with a amplified photodiode, as shown in Fig.~\ref{fig:TracesSUP}~b).    

\section{Countrate}
In comparison of the system detection efficiency (SDE) the single-pixel SNSPD was operated in the conventional and opto-electronic configuration. The repetition rate of the input laser was \SI{625}{k rep/s} with mean photon number of 0.83 for the conventional operation and \SI{600}{k rep/s} and a mean photon number of 1.1 for the opto-electronic operation. The dark countrate in the conventional operation reached a maximum of \SI{200}{cps}. The increase of dark count rate in the opto-electronic operation can be related to an increased electronic noise level.      

\begin{figure}[h]
    \centering
    \includegraphics[width=0.9\linewidth]{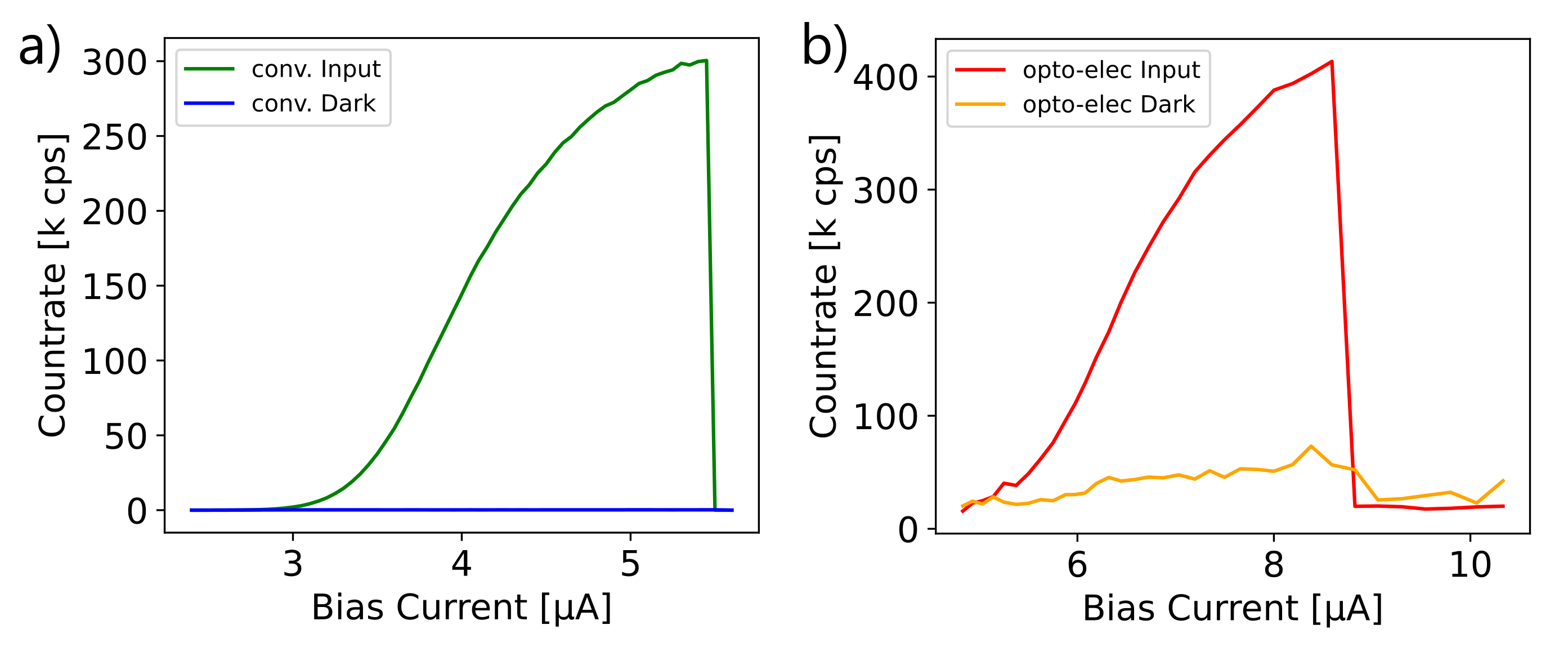}
    \caption{Acquired countrate in the SDE measurement with the single Pixel SNSPD a) Conventional current bias of the SNSPD with an input repetition rate of \SI{625}{k rep/s}. b) Opto-electronic operation of the SNSPD with the illumination of the cryogenic photodiode. The input repetition rate is \SI{600}{k rep/s}. }
    \label{fig:CountrateComparison}
\end{figure}

\end{document}